\documentclass[11pt]{article}

\usepackage{fullpage}
\usepackage[margin=1in]{geometry}
\usepackage{amsfonts,amsmath,latexsym,amssymb}
\usepackage[amsmath,amsthm,thmmarks]{ntheorem}
\usepackage{graphicx}
\usepackage[utf8]{inputenc}

\usepackage[ruled]{algorithm}
\usepackage{algorithmicx}
\usepackage[noend]{algpseudocode}
\floatname{algorithm}{Pseudocode}

\usepackage{enumerate}

\usepackage{scrextend}
\usepackage[T1]{fontenc}

\newcommand{\lastcorrections}%
{{
 \begin{sloppypar}
    \baselineskip -0.2in
    \tiny\bf\noindent
last corrections:\\
\end{sloppypar}
}}

\newcommand{\margincomment}[1]%
    {{%
      \marginpar{{\tiny\begin{minipage}{0.5in}
                       \begin{flushleft}
                          {#1}
                       \end{flushleft}
                       \end{minipage}
                }}
    }}

\newcommand{\ignore}[1]{}

\newcommand{\myparagraph}[1]{{\smallskip\noindent{\bf #1}}}
\newcommand{\emparagraph}[1]{{\smallskip\noindent\emph{#1}}}

\newcommand{\barS}{{\bar S}}

\newcommand{\calA}{{\cal A}}

\newcommand{\calH}{{\cal H}}

\newcommand{\calT}{{\cal T}}

\newcommand{\tildeS}{{\tilde{S}}}

\newcommand{\e}{\mathrm{e}}

\newcommand{\braced}[1]{{ \left\{ #1 \right\} }}

\newcommand{\ceiling}[1]{{ \lceil #1 \rceil }}
\newcommand{\floor}[1]{{ \lfloor #1 \rfloor }}

\newcommand{\suchthat}{{\;:\;}}
\newcommand{\assign}{\,{\leftarrow}\,}

\newcommand{\polylog}{{\mbox{\rm polylog}}}

\newcommand{\height}{{\textit{height}}}

\newcommand{\Prob}{{\mbox{\rm Pr}}}
\newcommand{\Exp}{{\mbox{\rm Exp}}}

\newcommand{\inv}[3]{($\textbf{#1}^{#3}_{#2}$)}

\newcommand{\vlabel}{{\textsf{label}}}

\newcommand{\RoundRobin}{{\textsc{RoundRobin}}}

\newcommand{\myTransmit}{{\textsc{Transmit}}}
\newcommand{\Hits}[3]{\textit{Hits}_{#1,#2}(#3)}

\newcommand{\algGatherOne}{{\textsc{SimpleGather}}}
\newcommand{\algGatherTwo}{{\textsc{FastGather}}}
\newcommand{\algGatherThree}{{\textsc{LinGather}}}

\newcommand{\attime}{{\mbox{\bf at time}}}

\newtheorem{theorem}{Theorem}

\newtheorem{lemma}{Lemma}

\newtheorem{claim}{Claim}

		{$\spadesuit$\smallskip}

\newenvironment{bigeqn*}{\large\begin{eqnarray*}}{\end{eqnarray*}}

\newcommand{\half}{{\textstyle\frac{1}{2}}}

\newcommand{\onefourth}{{\textstyle\frac{1}{4}}}

\begin{document}

\title{Faster Information Gathering in Ad-Hoc \\
		Radio Tree Networks}

\author{Marek Chrobak\thanks{%
			Department of Computer Science,
         	University of California at Riverside, USA.
			Research supported by NSF grants CCF-1217314 and CCF-1536026.
			}
		\and
		 Kevin Costello\thanks{%
			Department of Mathematics,
			University of California at Riverside, USA.
            Research supported by NSA grant H98230-13-1-0228
			}
		}

\maketitle

\begin{abstract}
We study information gathering in ad-hoc radio networks.
Initially, each node of the network has a piece of information called a \emph{rumor}, and the
overall objective is to gather all these rumors in the designated target node.
The ad-hoc property refers to the fact that the topology of the network is unknown
when the computation starts. Aggregation of rumors is not allowed, which means that
each node may transmit at most one rumor in one step.

We focus on networks with tree topologies, that is we assume that the network is a tree
with all edges directed towards the root, but, being ad-hoc, its actual topology is
not known.
We provide two deterministic algorithms for this problem. For the model that does not
assume any collision detection nor acknowledgement mechanisms, we
give an $O(n\log\log n)$-time algorithm, improving the previous upper
bound of $O(n\log n)$.
We also show that this running time can be further reduced to $O(n)$ if
the model allows for acknowledgements of successful transmissions.
\end{abstract}


\section{Introduction}
\label{sec: introduction}

We study the problem of \emph{information gathering} in ad-hoc radio networks.
Initially, each node of the network has a piece of information called a \emph{rumor}, and the
objective is to gather all these rumors, as quickly as possible, in the designated target node.
The nodes communicate by sending messages via radio transmissions. At any time step, several
nodes in the network may transmit. When a node
transmits a message, this message is sent immediately to all nodes within its range.
When two nodes send their messages to the same node at the same time, a
\emph{collision} occurs and neither message is received.
Aggregation of rumors is not allowed, which means that
each node may transmit at most one rumor in one step.

The network can be naturally modeled by a directed graph, where
an edge $(u,v)$ indicates that $v$ is in the range of $u$.
The ad-hoc property refers to the fact that the actual topology of the network is unknown
when the computation starts.
We assume that nodes are labeled by integers $0,1,...,n-1$.
An information gathering protocol determines a sequence of transmissions of a node,
based on its label and on the previously received messages.


\paragraph{Our results.}
In this paper, we focus on ad-hoc networks with tree topologies,
that is the underlying ad-hoc network is assumed
to be a tree with all edges directed towards the root, although the actual topology of this
tree is unknown.

We consider two variants of the problem. In the first one,
we do not assume any collision detection or acknowledgment
mechanisms, so none of the nodes (in particular neither the sender nor the intended recipient)
are notified about a collision after it occurred.
In this model, we give a deterministic algorithm that completes information gathering in time $O(n\log\log n)$.
Our result significantly improves the previous upper bound of $O(n\log n)$ from~\cite{ChrobakCGK_tree_gather_14}.
To our knowledge, no lower bound for this problem is known, apart from the
trivial bound of $\Omega(n)$ (since each rumor must be received by the root in a different time step).

In the second part of the paper, we also consider a variant where acknowledgments of
successful transmissions are provided to the sender. All the remaining nodes, though, including the
intended recipient, cannot distinguish between collisions and absence of transmissions.
Under this assumption, we show that the running time can be improved to $O(n)$, which is again optimal for trivial reasons, up to the implicit constant.

While we assume that all nodes are labelled $0,1,...,n-1$ (where $n$ is the number of vertices),
our algorithms' asymptotic running times remain the same if the labels are chosen from
a larger range $0,1,...,N-1$, as long as $N = O(n)$.


\paragraph{Related work.}
The problem of information gathering for trees was introduced in~\cite{ChrobakCGK_tree_gather_14},
where the model without any collision detection was studied. In addition to the
$O(n\log n)$-time algorithm without aggregation -- that we improve in this paper --
\cite{ChrobakCGK_tree_gather_14} develops an  $O(n)$-time algorithm for
the model with aggregation, where a message can include any number of rumors.
Another model studied in~\cite{ChrobakCGK_tree_gather_14}, called \emph{fire-and-forward}, requires
that a node cannot store any rumors; a rumor received by a node has to be either discarded
or immediately forwarded. For fire-and-forward protocols, a tight bound of $\Theta(n^{1.5})$ is given in~\cite{ChrobakCGK_tree_gather_14}.

The information gathering problem is closely related to two other information dissemination
primitives that have been well studied in the literature on
ad-hoc radio networks: broadcasting and gossiping. All the work discussed below
is for ad-hoc radio networks modeled by arbitrary directed graphs, and without
any collision detection capability.

In \emph{broadcasting}, a single rumor from a specified source node has to be delivered to all other
nodes in the network. The na\"{i}ve {\RoundRobin} algorithm (see the next section)
completes broadcasting in time $O(n^2)$. Following a sequence of
papers~\cite{Chrobak_etal_fast_02,Kowalski_Pelc_faster_04,Bruschi_etal_lower_bound_97,Chlebus_etal_broadcast_02,DeMarco_08,Czumaj_Rytter_broadcast_06}
where this na\"{i}ve bound was gradually improved,
it is now known that broadcasting can be solved  in time $O(n\log D\log\log (D\Delta/n))$~\cite{Czumaj_Davis_15b},
where $D$ is the diameter of $G$ and $\Delta$ is its maximum in-degree.
This nearly matches the  lower bound of $\Omega(n\log D)$ from~\cite{Clementi_etal_distributed_02}.
Randomized algorithms for broadcasting have also been
well studied \cite{alon_etal_lower_bound_91,Kushilevitz_Mansour_lower_bound_98,Czumaj_Rytter_broadcast_06}.

The \emph{gossiping} problem is an extension of broadcasting, where each node starts with its own
rumor, and all rumors need to be delivered to all nodes in the network.
The time complexity of deterministic algorithms for gossiping is a major open problem in the theory of
ad-hoc radio networks. Obviously, the lower bound of $\Omega(n\log D)$  for
broadcasting~\cite{Clementi_etal_distributed_02}
applies to gossiping as well, but no better lower bound is known.
It is also not known whether gossiping can be solved in time $O(n\,\polylog(n))$
with a deterministic algorithm, even if message aggregation is allowed.
The best currently known upper bound is
$O(n^{4/3}\log^4n)$ \cite{Gasieniec_etal_det_gossip_04}
(see~\cite{Chrobak_etal_fast_02,Xu_det_gossip_03} for some earlier work). The case when
no aggregation is allowed (or with limited aggregation) was studied in~\cite{ChristerssonGL_gossip_bounded_02}.
Randomized algorithms for gossiping have also been well
studied~\cite{Czumaj_Rytter_broadcast_06,Liu_Prabhakaran_randomized_02,Chrobak_etal_randomized_04}.
Interested readers can find more information about gossiping in the survey  paper \cite{Gasieniec_survey_gossiping_09}.


\paragraph{Connections to other problems.}
This research, as well as the earlier work in~\cite{ChrobakCGK_tree_gather_14},
was motivated by the connections between information gathering in trees and other
problems in distributed computing involving shared channels, including
gossiping in radio networks and MAC contention resolution.

For arbitrary graphs, assuming aggregation, one can solve the gossiping problem by running an
algorithm for information gathering and then broadcasting all rumors (as one message) to
all nodes in the network. Thus an $O(n\,\polylog(n))$-time algorithm for
information gathering would resolve in positive the earlier-discussed open question about the complexity of
gossiping. Due to this connection, developing an $O(n\,\polylog(n))$-time algorithm for information
gathering on arbitrary graphs is likely to be very difficult -- if possible at all.
We hope that developing efficient algorithms for trees, or for some other natural special cases,
will ultimately lead to some insights helpful in resolving the complexity of the gossiping problem
in arbitrary graphs.

Some algorithms for ad-hoc radio networks (see~\cite{ChristerssonGL_gossip_bounded_02,kowalski_selection_05}, for example)
involve constructing a spanning subtree of the network and disseminating information along this subtree.
Better algorithms for information gathering on trees may thus be useful in addressing problems
for arbitrary graphs.

The problem of \emph{contention resolution} for multiple-access channels (MAC) has been widely
studied in the literature. (See, for example,
\cite{DeMarco_Kowalski_13,demarco_kowalski_conflict_15,Anta_etal_13} and the references therein.)
There are in fact myriad of variants of this problem, depending on the characteristics of the
communication model. Generally, the instance of the MAC contention resolution problem involves
a collection of transmitters connected to a shared channel (e.g. ethernet). Some of these
transmitters need to send their messages across the channel, and the objective is to design
a distributed protocol that will allow them to do that.
The information gathering problem for trees is in essence an extension of MAC contention resolution
to multi-level hierarchies of channels, where transmitters have unique identifiers, and the
structure of this hierarchy is not known.


\section{Preliminaries}
\label{sec: preliminaries}

We now provide a formal definition of our model and introduce notation,
terminology, and some basic properties used throughout the paper.


\paragraph{Radio networks with tree topology.}
In the paper we focus exclusively on radio networks with tree topologies.
Such a network will be represented by a tree $\calT$ with root $r$ and with $n = |\calT|$ nodes.
The edges in $\calT$ are directed towards the root, representing
the direction of information flow: a node can send messages to its
parent, but not to its children.
We assume that each node $v\in \calT$ is assigned a unique label from
$[n] = \braced{0,1,...,n-1}$, and we denote this label by $\vlabel(v)$.

For a node $v$, by $\deg(v)$ we denote the \emph{degree} of $v$, which is
the number of $v$'s children. For any subtree $X$ of $\calT$ and a node
$v\in X$, we denote by $X_v$ the subtree of $X$ rooted at $v$ that consists
of all descendants of $v$ in $X$.

For any integer $\gamma = 1,2,...,n-1$ and any node $v$ of $\calT$
define the \emph{$\gamma$-height} of  $v$ as follows. If $v$ is a leaf then the
$\gamma$-height of $v$ is $0$.
If $v$ is an internal node then let $g$ be the maximum $\gamma$-height of a child
of $v$. If $v$ has fewer than $\gamma$ children of $\gamma$-height equal $g$ then
the $\gamma$-height of $v$ is $g$. Otherwise, the $\gamma$-height of $v$ is $g+1$.
The $\gamma$-height of $v$ will be denoted by $\height_\gamma(v)$.
In case when more than one tree are under consideration, to resolve potential
ambiguity we will write $\height_\gamma(v,\calT)$ for the $\gamma$-height of $v$ in $\calT$.
The $\gamma$-height of a tree $\calT$, denoted $\height_\gamma(\calT)$, is defined as $\height_\gamma(r)$,
that is the $\gamma$-height of its root.

\begin{figure}[ht]
\begin{center}
\includegraphics[width=3in]{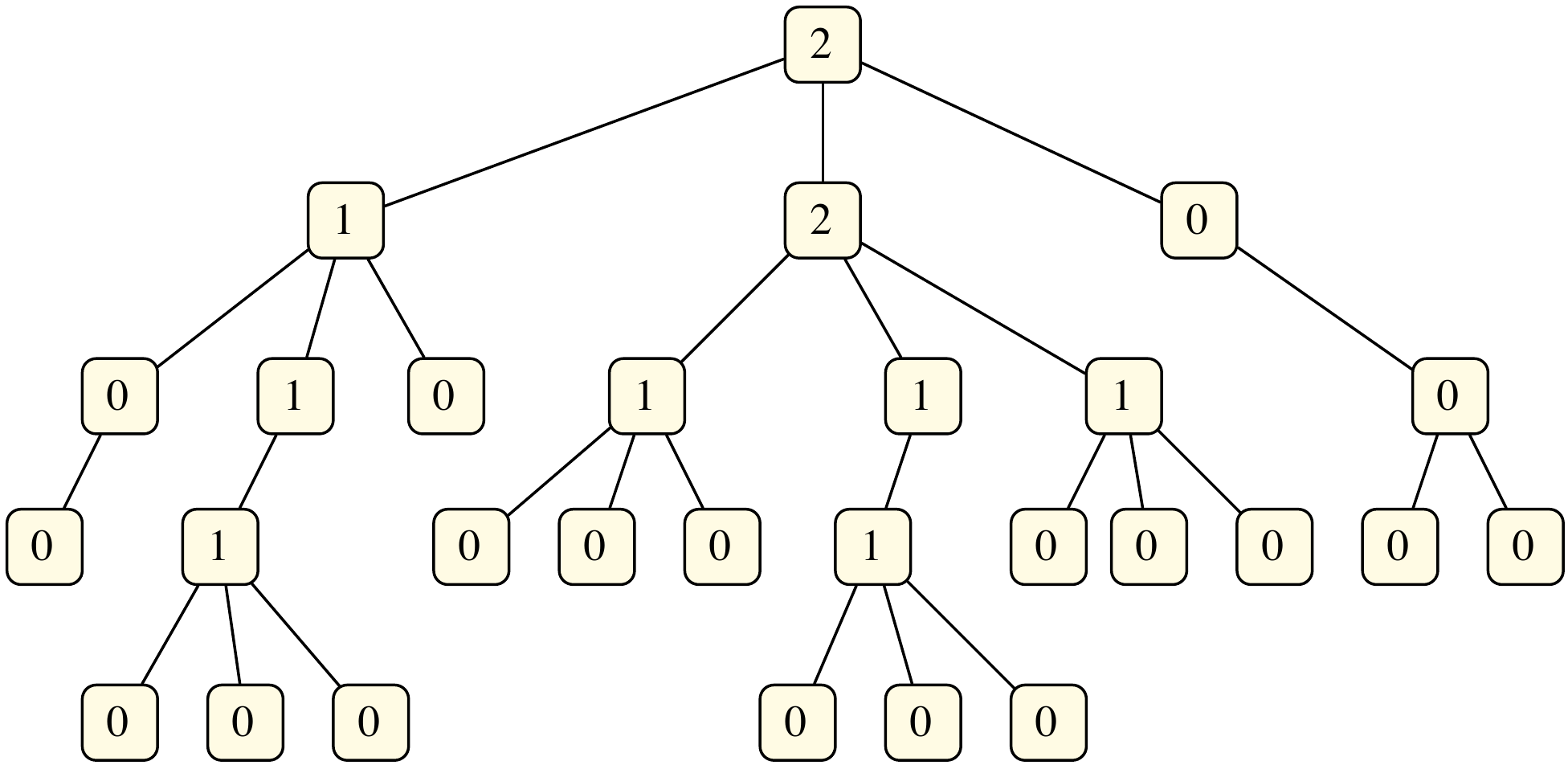}
\caption{An example showing a tree and the values of $3$-heights for all its nodes.}
\label{fig: gamma height example}
\end{center}
\end{figure}

Its name notwithstanding, the definition of $\gamma$-height is meant to capture the
``bushiness'' of a tree. For example, if $\calT$ is just a path
then its $\gamma$-height is equal $0$ for each $\gamma\ge 2$.
The concept of $\gamma$-height generalizes Strahler numbers~\cite{Strahler_52,Viennot_02},
introduced in hydrology to measure the size of streams in terms of the complexity
of their tributaries.
Figure~\ref{fig: gamma height example} gives an example of a tree and values of
$3$-heights for all its nodes.

The lemma below is a slight refinement of an analogous lemma
in~\cite{ChrobakCGK_tree_gather_14}, and it will play a critical role
in our algorithms.


\begin{lemma}\label{lem: gamma-heights}
Suppose that $\calT$ has $q$ leaves, and
let $2\le \gamma \le q$. Then
$\height_\gamma(\calT) \le \log_\gamma q$.
\end{lemma}

Equivalently, any tree having height $j$ must have at least $\gamma^j$ leaves.  This can be seen by induction on $j$ -- if $v$ is a vertex which is furthest from the room among all vertices of height $j$, then $v$ by definition has $\gamma$ descendants of height $j-1$, each of which has $\gamma^{j-1}$ leaf descendants by inductive hypothesis.


\paragraph{Information gathering protocols.}
Each node $v$ of $\calT$ has a label (or an identifier) associated with it, and
denoted $\vlabel(v)$.
When the computation is about to start, each node $v$ has also a
piece of information, $\rho_v$, that we call a \emph{rumor}.
The computation proceeds in discrete, synchronized time steps, numbered $0,1,2,...$.
At any step, $v$ can either be in the \emph{receiving state}, when it listens
to radio transmissions from other nodes, or in the \emph{transmitting state},
when it is allowed to transmit.
When $v$ transmits at a time $t$, the message from $v$ is sent immediately to its parent in $\calT$.
As we do not allow rumor aggregation, this message may contain at most one rumor, plus possibly
$O(\log n)$ bits of other information.
If $w$ is $v$'s parent, $w$ will receive $v$'s message if and only if $w$ is in the receiving state and
no collision occurred, that is if no other child of $w$ transmitted at time $t$.
In Sections~\ref{sec: simple algorithm} and~\ref{sec: nloglogn algorithm} we do not assume
any collision detection nor acknowledgement mechanisms, so if $v$'s
message collides with a message from one of its siblings, neither $v$ nor $w$ receive
any notification. (In other words, $w$ cannot distinguish collisions between its children's transmissions
from background noise.) We relax this requirement in Section~\ref{sec: linear with acknowledgments},
by assuming that $v$ (and only $v$) will obtain an acknowledgment from $w$ after each successful transmission.

The objective of an information gathering protocol is to deliver all rumors from $\calT$ to
its root $r$, as quickly as possible. Such a protocol needs to achieve its goal even
without the knowledge of the topology of $\calT$.
More formally, a gathering protocol $\calA$ can be defined as a function
that, at each time $t$, and for each given node $v$, determines the action of $v$
at time $t$ based only on $v$'s label and the information received by $v$ up to time $t$.
The action of $v$ at each time step $t$ involves choosing its state (either receiving or transmitting) and, if it is in the transmitting
state, choosing which rumor to transmit.

We will say that $\calA$ runs in time $T(n)$ if, for any tree $\calT$ and any
assignment of labels to its nodes, after at most $T(n)$ steps all rumors are delivered to $r$.

A simple example of an information gathering protocol is
called $\RoundRobin$. In $\RoundRobin$ nodes transmit one at a time, in $n$ rounds, where
in each round they transmit in the order $0,1,...,n-1$ of their labels.
For any node $v$, when it is its turn to transmit, $v$ transmits any
rumor from the set of rumors that have been
received so far (including its own rumor) but not yet transmitted.
In each round, each rumor that is still not in $r$ will get closer to $r$,
so after $n^2$ steps all rumors will reach $r$.


\paragraph{Strong $k$-selectors.}
Let $\barS = (S_1, S_2, ..., S_m)$ be a family of subsets of $\braced{0,1,...,n-1}$.
$\barS$ is called a
\emph{strong $k$-selector} if, for each $k$-element set $A\subseteq\braced{0,1,...,n-1}$
and each $a\in A$, there is a set $S_i$ such that $S_i\cap A = \braced{a}$.
As shown in~\cite{Erdos_etal_families_85,Clementi_etal_distributed_02},
for each $k$ there exists a strong $k$-selector
$\barS = (S_0, S_1, ..., S_{m-1})$ with $m = O(k^2\log n)$.
We will make extensive use of strong $k$-selectors in our algorithm.
At a certain time in the computation our protocols will ``run'' $\barS$, for an appropriate choice of $k$,
by which we mean that it will execute a sequence of $m$ consecutive steps,
such that in the $j$th step the nodes from $S_j$ will transmit, while those not in $S_j$
will stay quiet.
This will guarantee that, for any node $v$ with at most $k-1$ siblings,
there will be at least one step in the execution of $\barS$ where $v$ will transmit
but none of its siblings will. Therefore at least one of $v$'s transmissions will
be successful.


\section{An $O(n\sqrt{\log n})$-Time Protocol}
\label{sec: simple algorithm}

We first give a gathering protocol~{\algGatherOne}
for trees with running time $O(n\sqrt{\log n})$. Our faster
protocol will be presented in the next section. We fix three parameters:
$$K = 2^{\floor{\sqrt{\log n}}}, \, D = \ceiling{\log_Kn} = O(\sqrt{\log n}), \, D' = \ceiling{\log K^3} = O(\sqrt{\log n}).$$
We also fix a strong $K$-selector
$\barS = (S_0, S_1, ... , S_{m-1})$, where $m \le CK^2\log n$, for some integer constant $C$.

By Lemma~\ref{lem: gamma-heights}, we have that $\height_K(\calT) \le D$.
We call a node $v$ of $\calT$
\emph{light} if $|\calT_v|\le n/K^3$; otherwise we say that $v$ is \emph{heavy}.
Let $\calT'$ be the subtree of $\calT$ induced by the heavy nodes.
By the definition of heavy nodes, $\calT'$ has at most $K^3$ leaves,
so $\height_2(\calT')\le  D'$. Also, obviously, $r\in\calT'$.

To streamline the description of our algorithm
we will allow each node to receive and transmit messages at the
same time. We will also assume a preprocessing step allowing each $v$ to know both the
size of its subtree $\calT_v$ (in particular, whether it is in $\calT'$ or not), its $K$-height, and, if it is in $\calT'$, its $2$-height in the subtree $\calT'$.
We later explain both the preprocessing and how to modify the algorithm to remove the receive/transmit assumption.

The algorithm consists of two epochs. Epoch~$1$ consists of $D+1$
stages, each lasting $O(n)$ steps. In this epoch only light vertices participate.
The purpose of this epoch is to gather all rumors from $\calT$ in $\calT'$.
Epoch~2 has $D'+1$ stages and only heavy vertices participate in
the computation during this epoch. The purpose of epoch~2 is to
deliver all rumors from $\calT'$ to $r$.
We describe the computation in the two epochs separately.

A detailed description of Algorithm~{\algGatherOne} is given in Pseudocode~\ref{alg: gather1}.
To distinguish between computation steps (which do not consume time) and communication steps, we use
command ``{{\attime} $t$}''. When the algorithm reaches this command it
waits until time step $t$ to continue processing. Each message transmission takes one time step.
For each node $v$ we maintain a set $B_v$ of rumors received by $v$, including its own
rumor $\rho_v$.


\medskip
\emparagraph{Epoch~1: light vertices.}
Let $0 \le h \le D$, and let $v$ be a light vertex whose $K$-height equals $h$.
Then $v$ will be active only during stage $h$ which starts at time $\alpha_h = (C+1)hn$. This stage
is divided into two parts.

In the first part of stage $h$,
 $v$ will transmit according to the strong $K$-selector $\barS$.
Specifically, this part has $n/K^3$ iterations, where each iteration corresponds to a
complete execution of $\barS$. At any time,
some of the rumors in $B_v$ may be marked; the marking on a rumor indicates that the algorithm has
already attempted to transmit it using $\barS$.
At the beginning of each iteration, $v$ chooses any rumor $\rho_z\in B_v$ it has not yet marked, then transmits
$\rho_z$ in the steps that use sets $S_i$ containing the label of $v$.
This $\rho_z$ is then marked.
If the parent $u$ of $v$ has degree at most $K$, the definition of
strong $K$-selectors guarantees that $\rho_z$ will be received by $u$, but if $u$'s
degree is larger it may not have received $\rho_z$.
Note that the total number of steps required for this part of
stage $h$ is $(n/K^3)\cdot m \le Cn$, so these steps will be completed
before the second part of stage $h$ starts.

In the second part, that starts at time $\alpha_h + Cn$,
we simply run a variant of the $\RoundRobin$ protocol, but cycling through rumors instead of nodes: in the $l$-th step of this part, all nodes  holding the rumor of the node with label $l$ transmit that rumor (note that due to the tree topology it is impossible for two siblings to both be holding rumor $\ell$).

\smallskip

We claim that the following invariant holds for all $h = 0,1,...,D$:
\begin{description}
	\item{\inv{I}{h}{}} Let $w\in\calT$ and let $u$ be a light child of $w$ with $\height_K(u) \le h-1$.
	Then at time $\alpha_{h}$ node $w$ has received all rumors from $\calT_u$.
\end{description}
To prove this invariant we proceed by induction on $h$. If $h=0$ the invariant {\inv{I}{0}{}} holds
vacuously. So suppose that invariant~{\inv{I}{h}{}} holds for some value of $h$. We want to prove
that {\inv{I}{h+1}{}} is true when stage $h+1$ starts.
We thus need to prove the following claim:
if $u$ is a light child of $w$ with $\height_K(u) \le h$
then at time $\alpha_{h+1}$ all rumors from $\calT_u$ will arrive in $w$.

If  $\height_K(u) \le h-1$ then the claim holds, immediately from the inductive assumption {\inv{I}{h}{}}.
So assume that $\height_K(u) = h$. Consider the subtree $H$ rooted at $u$
and containing all descendants of $u$ whose $K$-height is equal to $h$.
By the inductive assumption, at time $\alpha_h$ any $w'\in H$ has all rumors from the subtrees rooted at
its descendants of $K$-height smaller than $h$, in addition to its own rumor $\rho_{w'}$.
Therefore all rumors from $\calT_u$ are already in $H$ and each of them has exactly one copy in $H$,
because all nodes in $H$ were idle before time $\alpha_h$.

When the algorithm executes the first part of stage $h$ on $H$, then each $v$ node in $H$ whose parent is also in $H$ will successfully transmit an unmarked rumor
during each pass through the strong $K$-selector -- indeed, our definition of $H$ guarantees that $v$ has at most $K-1$ siblings in $H$, so by the definition of strong selector it must succeed at least once.  We make the following additional claim:

\begin{claim}
At all times during stage $h$, the collection of nodes in $H$ still holding unmarked rumors forms an induced tree of $H$
\end{claim}

The claim follows from induction: At the beginning of the stage the nodes in $H$ still hold their own original rumor, and it is unmarked since those nodes were idle so far.  As the stage progresses, each parent of a transmitting child will receive a new (and therefore not yet marked) rumor during each run through the strong selector, so no holes can ever form.

In particular, node $u$ will receive a new rumor during every run through the strong selector until it has received all rumors from its subtree.  Since the tree originally held at most $|\calT_u|\le n/K^3$ rumors originally, $u$ must have received all rumors from its subtree after at most $n/K^3$ runs through the selector.

Note that,
as $\height_K(u) =h$, $u$ will also attempt to transmit its rumors to $w$ during this part, but,
since we are not making any assumptions about the degree of $w$, there is no guarantee that $w$ will receive them.
This is where the second part of this stage is needed. Since in the second part each
rumor is transmitted without collisions, all rumors from $u$ will reach $w$ before
time $\alpha_{h+1}$, completing the inductive step and the proof that {\inv{I}{h+1}{}} holds.

\begin{figure}[ht]
\begin{center}
\includegraphics[width=2.75in]{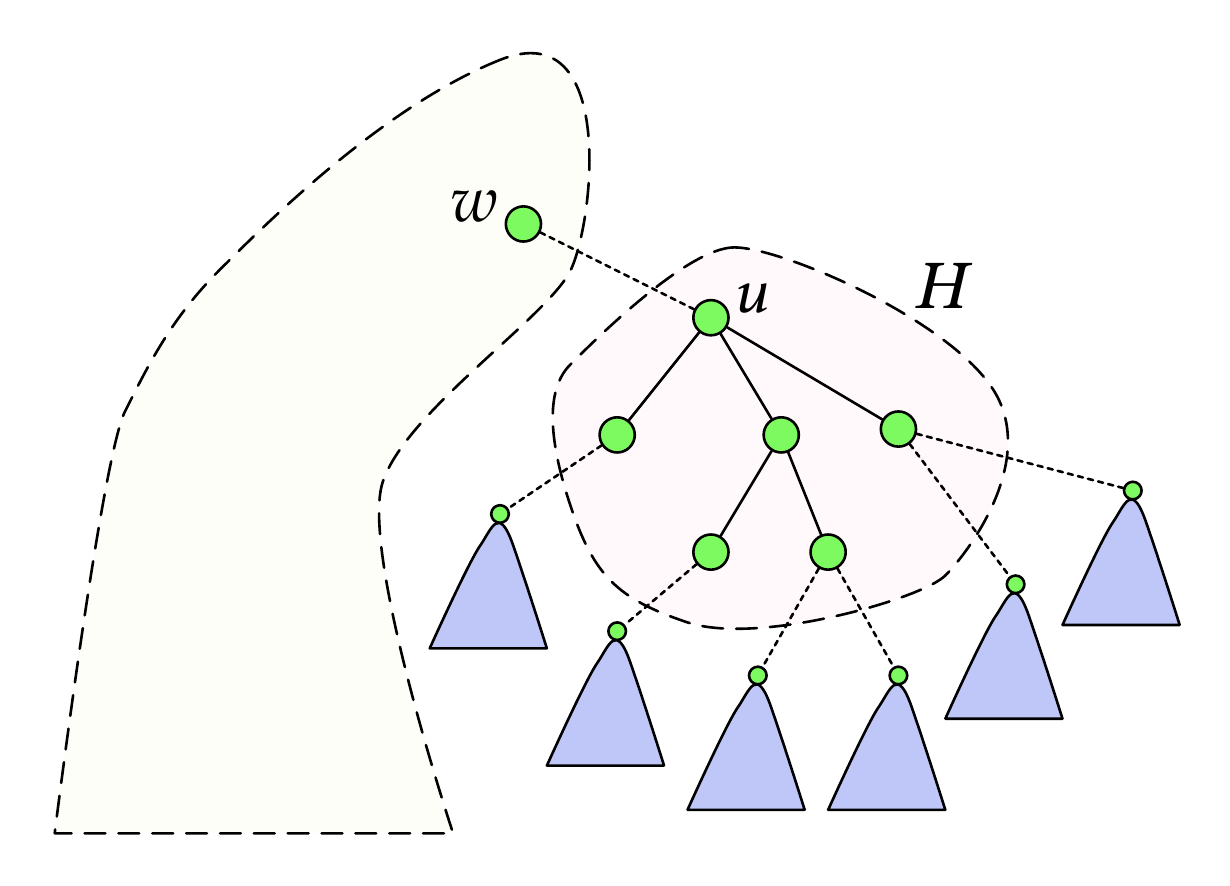}
\caption{Proving Invariant~{\inv{I}{h}{}}. Dark-shaded subtrees of $\calT_u$ consist of light nodes with
	$K$-height at most $h-1$. $H$ consists of the descendants of $u$ with $K$-height equal $h$.}
\label{fig: invariant I1}
\end{center}
\end{figure}

In particular, using Invariant~{\inv{I}{h}{}} for $h=D$, we obtain that after epoch~1
each heavy node $w$ will have received rumors from the subtrees rooted at all its light
children. Therefore at that time all rumors from $\calT$ will be already in $\calT'$,
with each rumor having exactly one copy in $\calT'$.


\newcommand{\LineIf}[2]{\State \algorithmicif\ {#1}\ \algorithmicthen\ {#2} }
\newcommand{\LineFor}[2]{\State \algorithmicfor\ {#1}\ \algorithmicdo\ {#2} }

\newcommand{\myComment}[1]{\Comment {\makebox[2in]{#1\hfill}}}
\begin{algorithm}
  \caption{$\algGatherOne(v)$}
  \label{alg: gather1}
  \begin{algorithmic}[1]
	\State $K = 2^{\floor{\sqrt{\log n}}}$, $D = \ceiling{\log_Kn}$
	\State $B_v \assign \braced{\rho_v}$
	\myComment{Initially $v$ has only $\rho_v$}
	\State \textbf{Throughout:} all rumors received by $v$ are automatically added to $B_v$
	\If {$|T_v|\le n/K^3$}
		\myComment {$v$ is light (epoch~1)}
		\State $h\assign \height_K(v,\calT)$ ; \ $\alpha_h \assign (C+1)nh$
		\myComment {$v$ participates in stage $h$}
		\For {$i = 0,1,...,n/K^3-1$}
		    \myComment {iteration $i$}
		    \State {\attime} $\alpha_h + im$
			\If {$B_v$ contains an unmarked rumor}
				\myComment{Part 1: strong $K$-selector}			
				\State choose any unmarked $\rho_z\in B_v$ and mark it
				\For {$j=0,1,...,m-1$}
					\State {\attime} $\alpha_h + im+j$
					\LineIf {$\vlabel(v)\in S_j$}{$\myTransmit(\rho_z)$}
				\EndFor
			\EndIf
		\EndFor
		\For{$l = 0,1,...,n-1$}
			\myComment{Part 2: RoundRobin}
			\State {\attime} $\alpha_h + Cn + l$
			\State $z\assign \textrm{node with $\vlabel(z) = l$}$
			\LineIf {$\rho_z\in B_v$}{$\myTransmit(\rho_z)$}
		\EndFor
	\Else
		\myComment{$v$ is heavy (epoch~2)}
		\State $g\assign \height_2(v,\calT')$ ; \ $\alpha'_g \assign \alpha_{D+1} + 2ng$
		\myComment {$v$ participates in stage $g$}
		\For {$i = 0,1,...,n-1$}
		    \myComment{Part 1: all nodes transmit}
			\State {\attime} $\alpha'_g+i$
			\If {$B_v$ contains an unmarked rumor}			
				\State choose any unmarked $\rho_z\in B_v$ and mark it
				\State $\myTransmit(\rho_z)$
			\EndIf
		\EndFor
		\For {$l = 0,1,...,n-1$}
			\myComment{Part 2: RoundRobin}
			\State {\attime} $\alpha'_g + 2n + l$
			\State $z\assign \textrm{node with $\vlabel(z) = l$}$
			\LineIf {$\rho_z\in B_v$}{$\myTransmit(\rho_z)$}
		\EndFor
	\EndIf
  \end{algorithmic}
\end{algorithm}


\medskip
\emparagraph{Epoch~2: heavy vertices.}
In this epoch we have at most $D'+1$ stages, and only heavy nodes in $\calT'$
participate in the computation. When the epoch starts, all rumors are already in $\calT'$.
In stage $D+1+g$ the nodes in $\calT'$ whose $2$-height is equal $g$ will participate.
Similar to the stages of epoch~1, this stage has two parts and the second part
executes $\RoundRobin$, as before. The difference is that now, in
the first part, instead of using the strong $K$-selector, each heavy node will
transmit at each of the $n$ steps.

\smallskip

We need to show that $r$ will receive all rumors at the end.
The argument is similar as for light vertices, but with a twist, since we do not
use selectors now; instead we have steps when all nodes transmit. In essence, we show that each stage
reduces by at least one the $2$-depth of the minimum subtree of $\calT'$ that contains all rumors.

Specifically, we show that the following invariant holds for all $g = 0,1,...,D'$:
\begin{description}
	\item{\inv{J}{g}{}} Let $w\in\calT'$ and let $u\in\calT'$ be a child of $w$ with $\height_2(u,\calT') \le g-1$.
	Then at time $\alpha'_{g}$ node $w$ has received all rumors from $\calT_u$.
\end{description}
We prove invariant {\inv{J}{g}{}} by induction on $g$. For $g=0$, {\inv{J}{0}{}} holds vacuously.
Assume that {\inv{J}{g}{}} holds for some $g$. We claim that {\inv{J}{g+1}{}} holds right after stage $g$.

Choose any child $u$ of $w$ with $\height_2(u,\calT') \le g$. If
$\height_2(u,\calT') \le g-1$, we are done, by the inductive assumption.
So we can assume that $\height_2(u,\calT') = g$.
Let $P$ be the subtree of $\calT'$ rooted at $u$ and consisting of all
descendants of $u$ whose $2$-height in $\calT'$ is equal $g$.
Then $P$ is simply a path. By the inductive assumption, for each $w'\in P$,
all rumors from the subtrees of $w'$ rooted at its children of $2$-height at most
$g-1$ are in $w'$. Thus all rumors from $\calT_u$ are already in $P$.
All nodes in $P$ participate in stage $g$, but their
children outside $P$ do not transmit. Therefore each transmission
from any node $x\in P-\braced{u}$ during stage $g$ will be successful.
Due to pipelining, all rumors from $P$ will reach $u$ after the first part of
stage $g$. (This conclusion can also be derived from treating this computation on $P$
as an instance of the token collection game in Section~\ref{sec: preliminaries},
with each step of the transmissions being one step of the game.)
In the second part, all rumors from $u$ will be successfully sent to $w$.
So after stage $g$ all rumors from $\calT_u$ will be in $w$, completing
the proof that {\inv{J}{g+1}{}} holds.


\paragraph{Removing simplifying assumptions.}
At the beginning of this section we made some simplifying assumptions. It still remains to explain how to
modify our algorithm so that it works even if these assumptions do not hold.
These modification are similar to those described in~\cite{ChrobakCGK_tree_gather_14}, but we
include them here for the sake of completeness.

First, we assumed a preprocessing step whereby each $v$ knows certain parameters of its subtree $\calT_v$,
including the size, its $K$-height, etc.  The justification for this lies in the algorithm from~\cite{ChrobakCGK_tree_gather_14} for information gathering in trees with aggregation.  Such an algorithm can be modified to compute in linear time any function $f$ such that $f(v)$ is uniquely determined by the values of $f$ on the children of $v$.  The modification is that each node $u$, when its sends its message (which, in the
algorithm from~\cite{ChrobakCGK_tree_gather_14} contains all rumors from $\calT_u$), it will
instead send the value of $f(u)$. A node $v$, after it receives all values of $f(u)$ from each child $u$,
will then compute $f(v)$\footnote{It needs to be emphasized here that in our model only communication
steps contribute to the running time; all calculations are assumed to be instantaneous.}.

We also assumed that each node can receive and transmit messages at the same time. We now need
to modify the algorithm so that it receives messages only in the receiving state and transmits
only in the transmitting state. For the {\RoundRobin} steps this is trivial: a node $v$
is in the transmitting state only if it is scheduled to transmit, otherwise it is in the receiving state.
For other steps, we will explain the modification for light and heavy nodes separately.

Consider the computation of the light nodes during the steps when they transmit according to
the strong selector. Instead of the strong $K$-selector, we can use the strong $(K+1)$-selector, which will
not affect the asymptotic running time. When a node $v$ is scheduled to transmit, it enters the
transmitting state, otherwise it is in the receiving state.
In the proof, where we argue that the message from $v$ will reach its parent, instead of applying the selector
argument to $v$ and its siblings, we apply it to the set of nodes consisting of $v$, its siblings,
and its parent, arguing that there will be a step when $v$ is the only node transmitting among
its siblings and its parent is in the receiving state.

Finally, consider the computation of the heavy nodes, at steps when all of them transmit.
We modify the algorithm so that, in any stage $g$, the iteration (in Line~18) of these steps is preceded by
$O(n)$-time preprocessing. Recall that the nodes whose $2$-height in $\calT'$ is equal $g$ form
disjoint paths. We can run a one round of {\RoundRobin} where each node transmits an arbitrary message.
This way, each node will know whether it is the first node on one of these paths or not.
If a node $x$ is first on some path, say $P$, $x$ sends a message along this path, so that each
node $y\in P$ can compute its distance from $x$.
Then, in the part where all nodes transmit, we replace each step by two consecutive steps (even and odd), and
we use parity to synchronize the
computation along these paths: the nodes at even positions are in the receiving state at even steps
and in the transmitting state at odd steps, and the nodes at odd positions do the opposite.


\smallskip

Summarizing this section, we have presented Algorithm~{\algGatherOne} that
completes information gathering in ad-hoc radio networks with tree topologies in
time $O(n\sqrt{\log n})$. In the next section, we will show how to improve
this bound to $O(n\log\log n)$.


\section{A Protocol with Running Time $O(n\log\log n)$}
\label{sec: nloglogn algorithm}

In this section we consider the same basic model of information gathering in trees
as in Section~\ref{sec: simple algorithm}, that is, the model does not provide
any collision detection and rumor aggregation is not allowed.
We show that how to refine our $O(n\sqrt{\log n})$ protocol~{\algGatherOne}
to improve the running time to $O(n\log\log n)$.
This is the first main result of our paper, as summarized in the theorem below.


\begin{theorem}\label{thm: nloglogn algorithm}
The problem of information gathering on trees, without rumor aggregation, can be
solved in time $O(n\log\log n)$.	
\end{theorem}


Our protocol that achieves running time $O(n\log\log n)$ will be called ~{\algGatherTwo}.
This protocol can be thought of as an iterative application of the idea
behind Algorithm~{\algGatherOne} from Section~\ref{sec: simple algorithm}.
We assume that the reader is familiar with Algorithm~{\algGatherOne} and its analysis,
and in our presentation
we will focus on the high level ideas behind Algorithm~{\algGatherTwo}, referring
the reader to Section~\ref{sec: simple algorithm} for the implementation of some details.

As before, we use notation $\calT$ for the input tree and $n = |\calT|$
denotes the number of vertices in $\calT$. We assume that $n$ is sufficiently large, and we will establish
some needed lower bounds for $n$ as we work through the proof.
We fix some arbitrary integer constant $\beta \ge 2$.
For $\ell = 1,2, ...$, let $K_\ell = \ceiling{ n^{\beta^{-\ell}}}$.
So $K_1 = \ceiling{n^{1/\beta}}$, the sequence $(K_\ell)_\ell$ is non-increasing,
and $\lim_{\ell\to \infty} K_\ell = 2$.
Let $L$ be the largest value of $\ell$ for which $n^{\beta^{-\ell}} \ge \log n$.
(Note that $L$ is well defined for sufficiently large $n$, since $\beta$ is fixed). We thus have the following
exact and asymptotic bounds:
\begin{align*}
L &\le \log_\beta(\log n/\log\log n)
	&
	K_L &\ge \log n
	\\
L &= \Theta(\log\log n)
	&
	K_L &= \Theta(\log n).
\end{align*}
For $\ell = 1,2,...,L$, by $\barS^\ell = (S^\ell_1, S^\ell_2,...,S^\ell_{m_\ell})$
we denote a strong $K_\ell$-selector of size
$m_\ell \le C K_\ell^2\log n$, for some integer constant $C$. As discussed
in Section~\ref{sec: preliminaries}, such selectors $\barS^\ell$ exist.

Let $\calT^{(0)} = \calT$, and
for each $\ell = 1,2,...,L$, let $\calT^{(\ell)}$ be the subtree of
$\calT$ induced by the nodes $v$ with $|\calT_v| \ge n/K_{\ell}^3$.
Each tree $\calT^{(\ell)}$ is rooted at $r$, and
$\calT^{(\ell)} \subseteq \calT^{(\ell-1)}$ for $\ell \ge 1$.
For $\ell\neq\ 0$,
the definition of $\calT^{(\ell)}$ implies also that it has at most
$K_{\ell}^3$ leaves, so, by Lemma~\ref{lem: gamma-heights},
its $K_{\ell+1}$-height is at most $\log_{K_{\ell+1}}(K_{\ell}^3)$.
Since $K_\ell \le 2n^{\beta^{-\ell}}$ and $K_{\ell+1}\ge n^{\beta^{-(\ell+1)}}$,
we have
\begin{align*}
	\log_{K_{\ell+1}}(K_{\ell}^3)
		&= 3\, \log_{K_{\ell+1}}(K_{\ell})
		\\
		&\le  3\, \log_{ n^{\beta^{-(\ell+1)}} }(2n^{\beta^{-\ell}})
		\\
		&\le  3\, \log_{ n^{\beta^{-(\ell+1)}} }n^{\beta^{-\ell}}
				+ 3\, \log_{ n^{\beta^{-(\ell+1)}} } 2
		\\
		&= 3\beta + \frac{3\beta^{\ell+1}}{\log n}
		\le 3\beta + \frac{3\beta^{L+1}}{\log n}
		\le 3\beta+1,
\end{align*}
where the last inequality holds as long as $3\beta^{L+1}\le \log n$, which is
true if $n$ is large enough (since by the first of the "exact and asymptotic bounds" above we have $\beta^L \leq \log n/\log \log n$). We thus obtain that
the $K_{\ell+1}$-height of $\calT^{(\ell)}$ is at most $D = 3 \beta+1 = O(1)$.

\smallskip

As in the previous section, we will make some simplifying assumptions.
First, we will assume that all nodes can receive and transmit messages
at the same time. Second, we will also assume that each node $v$
knows the size of its subtree $|\calT_v|$ and
its $K_\ell$-heights, for each $\ell \le L$.
After completing the description of the algorithm we will explain
how to modify it so that these assumptions will not be needed.

Algorithm~{\algGatherTwo} consists of $L+1$ epochs, numbered $1,2,...,L+1$.
In each epoch $\ell \le L$ only the
nodes in $\calT^{(\ell-1)} -  \calT^{(\ell)}$ participate.
At the beginning of epoch $\ell$ all rumors will be already collected in $\calT^{(\ell-1)}$, and
the purpose of epoch $\ell$ is to move all these rumors into $\calT^{(\ell)}$.
Each of these $L$ epochs will run in time $O(n)$, so their total running time
will be $O(nL) = O(n\log\log n)$.
In epoch $L+1$, only the nodes in $\calT^{(L)}$ participate. At the beginning of
this epoch all rumors will be already in $\calT^{(L)}$, and when this epoch is
complete, all rumors will be collected in $r$. This epoch will take time $O(n\log\log n)$.
Thus the total running time will be also $O(n\log\log n)$. We now provide the details.


\medskip
\emparagraph{Epochs $\ell = 1,2,...,L$.}
In epoch~$\ell$, only the nodes in $\calT^{(\ell-1)} -  \calT^{(\ell)}$ are active, all other nodes will be quiet.
The computation in this epoch is very similar to the computation of light nodes (in epoch~$1$) in Algorithm~{\algGatherOne}.
Epoch~$\ell$ starts at time $\gamma_\ell = (D+1)(C+1)(\ell-1) n$ and lasts $(D+1)(C+1)n$ steps.

Let $v\in \calT^{(\ell-1)} -  \calT^{(\ell)}$. The computation of $v$ in
epoch~$\ell$ consists of $D+1$ identical stages.
Each stage~$h = 0,1,...,D$ starts at time step $\alpha_{\ell,h} = \gamma_\ell + (C+1)hn$ and lasts
$(C+1)n$ steps.

Stage $h$ has two parts.
The first part starts at time $\alpha_{\ell,h}$ and lasts time $Cn$.
During this part we execute $\ceiling{n/K_\ell^3}$ iterations, each iteration
consisting of running the strong $K_{\ell+1}$-selector $\barS^\ell$.
The time needed to execute these iterations is at most
\begin{align*}
	\ceiling{ \frac{n}{K_\ell^3} } \cdot CK^2_{\ell+1}\log n
		&\le Cn \cdot \frac{ 2 K^2_{\ell+1} \log n } { K_\ell^3 }
		\\
		&\le Cn \cdot \frac{ 8 n^{2\beta^{-(\ell+1)}} \log n } { n^{3\beta^{-\ell}} }
		\\
		&= Cn \cdot \frac{8 \log n}{n^{(3-2/\beta)\beta^{-\ell}}}
		\\
			&\le Cn \cdot \frac{8\log n}{n^{(3-2/\beta)\beta^{-L}}}
  	    \le  Cn,
\end{align*}
where the last inequality holds as long as $n^{3-2/\beta}\beta^{-L} \ge 8\log n$, which is
again true for sufficiently large $n$ (recall that $\beta \geq 2$ is constant, and $L=\Theta(\log \log n)$).

Thus all iterations executing the strong selector will complete before time $\alpha_{\ell,h}+Cn$.
Then $v$ stays idle until time $\alpha_{\ell,h}+Cn$,
which is when the second part starts. In the second part
we run the {\RoundRobin} protocol, which takes $n$ steps. So stage $h$ will
complete right before step $\alpha_{\ell,h}+(C+1)n = \alpha_{\ell,h+1}$.
Note that the whole epoch will last $(D+1)(C+1) n$ steps, as needed.

\smallskip

We claim that the algorithm preserves the following invariant for $\ell = 1,2,...,L$:
\begin{description}
\item{\inv{I}{}{\ell}}
Let $w\in\calT$ and let $u\in  \calT- \calT^{(\ell-1)}$ be a child of $w$. Then
$w$ will receive all rumors from $\calT_u$ before time $\gamma_\ell$, that is
before epoch $\ell$ starts.
\end{description}
For $\ell = 1$, invariant~{\inv{I}{}{1}} holds vacuously, because $\calT^{(0)} = \calT$.
In the inductive step, assume that {\inv{I}{}{\ell}} holds for some epoch $\ell$. We
want to show that {\inv{I}{}{\ell+1}} holds right after epoch $\ell$ ends. In other words,
we will show that if $w$ has a child $u\in  \calT- \calT^{(\ell)}$ then $w$ will
receive all rumors from $\calT_u$ before time $\gamma_{\ell+1}$.

So let $u \in \calT-\calT^{(\ell)}$. If $u \not\in\calT^{(\ell-1)}$ then
{\inv{I}{}{\ell+1}} holds for $u$, directly from the inductive assumption. We can thus assume
that $u\in \calT^{(\ell-1)} - \calT^{(\ell)}$.

The argument now is very similar to that for Algorithm~$\algGatherOne$
in Section~\ref{sec: simple algorithm}, when we analyzed epoch~1. For each
$h = 0,1,...,D$ we prove a refined version of condition~{\inv{I}{}{\ell+1}}:
\begin{description}
\item{{\inv{I}{h}{\ell+1}}}
Let $w\in\calT$ and let $u\in \calT^{(\ell-1)} - \calT^{(\ell)}$ be child of $w$ with
$\height_{K_\ell}(u,\calT^{(\ell-1)}) \le h-1$. Then
$w$ will receive all rumors from $\calT_u$ before time
$\alpha_{\ell,h}$, that is before stage $h$.
\end{description}
The proof is the same as for Invariant~{\inv{I}{h}{}} in Section~\ref{sec: simple algorithm},
proceeding by induction on $h$. For each fixed $h$ we consider a subtree
$\calH$ rooted at $u$ and consisting of all
descendants of $u$ in $\calT^{(\ell-1)}$ whose $K_\ell$-height is at least $h$.
By the inductive assumption,
at the beginning of stage $h$ all rumors from $\calT_u$ are already in $\calH$. Then,
the executions of $\barS^\ell$, followed by the execution of
{\RoundRobin}, will move all rumors from $\calH$ to $w$. We omit the details here.

By applying condition~{\inv{I}{h}{\ell+1}} with $h = D$, we obtain that
after all stages of epoch $\ell$ are complete, that is at right before time $\gamma_{\ell+1}$,
$w$ will receive all rumors from $\calT_u$.
Thus~$\ell$ invariant~{\inv{I}{}{\ell+1}} will hold.

\medskip
\emparagraph{Epoch $L+1$.}
Due to the definition of $L$, we have that $\calT^{(L)}$ contains at most $K^3_L = O(\log^3n)$
leaves, so its $2$-depth is at most $D' = \log(K^3_L)=O(\log\log n)$, by Lemma~\ref{lem: gamma-heights}.
The computation in this epoch is similar to epoch~2 from~Algorithm~{\algGatherOne}.
As before, this epoch consists of $D'+1$ stages, where
each stage $g = 0,1,...,D'$ has two parts. In the first part, we have $n$
steps in which each node transmits.
In the second part, also of length $n$, we run one iteration of {\RoundRobin}.

Let $\alpha'_g = \gamma_L + 2gn$.
To prove correctness, we show that the following invariant holds for all
stages $g = 0,1,...,D'$:
\begin{description}
	\item{\inv{J}{g}{}} Let $w\in\calT^{(L)}$ and let $u\in\calT^{(L)}$ be a child of $w$
	 with $\height_2(u,\calT^{(L)}) \le g-1$.
	Then at time $\alpha'_{g}$ node $w$ has received all rumors from $\calT_u$.
\end{description}
The proof is identical to the proof of the analogous
Invariant~{\inv{J}{g}{}} in Section~\ref{sec: simple algorithm}, so we omit it here.
Applying Invariant~{\inv{J}{g}{}} with $g=D'$, we conclude that
after stage $D'$, the root $r$ will receive all rumors.

\medskip
As for the running time, we recall that $L = O(\log \log n)$.
Each epoch $\ell = 1,2,...,L$ has $D+1 = O(1)$ stages, where each stage takes time $O(n)$,
so the execution of the first $L$ epochs will take time $O(n\log\log n)$.
Epoch $L+1$ has $D'+1 = O(\log\log n)$ stages, each stage consisting of
$O(n)$ steps, so this epoch will complete in time $O(n\log\log n)$.
We thus have that the overall running time of our protocol is $O(n\log\log n)$.

\medskip
It remains to explain that the simplifying assumptions we made at the
beginning of this section are not needed.
Computing all subtree sizes and all $K_\ell$-heights can be done
recursively bottom-up, using the linear-time information gathering
 algorithm from~\cite{ChrobakCGK_tree_gather_14} that uses aggregation.
This was explained in Section~\ref{sec: simple algorithm}.
The difference now is that each node has to compute $L+1 = O(\log\log n)$
values $K_\ell$, and, since we limit bookkeeping information in each
message up to $O(\log n)$ bits, these values need to be computed
separately. Nevertheless, the total pre-computation time will still be $O(n\log\log n)$.

Removing the assumption that nodes can receive and transmit at the same
time can be done in the same way as in Section~\ref{sec: simple algorithm}.
Roughly, in each epoch~$\ell = 1,2,...,L$, any node $v \in \calT^{(\ell-1)} -  \calT^{(\ell)}$
uses a strong $(K_\ell+1)$-selector (instead of a strong $K_\ell$-selector)
to determine whether to be in the receiving or transmitting state.
In epoch~$L$ the computation (in the steps when all nodes transmit)
is synchronized by transmitting a control message along induced paths,
and then choosing the receiving or transmitting state according to node parity.

\medskip
Summarizing, we have proved that the running time of
Algorithm~$\algGatherOne$ is $O(n\log\log n)$, thus completing the
proof of Theorem~\ref{thm: nloglogn algorithm}.


\section{An $O(n)$-time Protocol with Acknowledgments}
\label{sec: linear with acknowledgments}

In this section we consider a slightly different communication model
from that in Sections~\ref{sec: simple algorithm} and~\ref{sec: nloglogn algorithm}.
We now assume that acknowledgments of successful transmissions are provided to the sender.
All the remaining nodes, including the
intended recipient, cannot distinguish between collisions and absence of transmissions.
The main result of this section, as summarized in the theorem below, is that
with this feature it is possible to achieve the optimal time $O(n)$.


\begin{theorem}\label{thm: linear algorithm}
The problem of information gathering on trees without rumor aggregation can be
solved in time $O(n)$ if acknowledgments are provided.  	
\end{theorem}


The overall structure of our $O(n)$ time protocol, called Algorithm~{\algGatherThree},
is similar to Algorithm~{\algGatherOne}. It consists of two epochs.
The first epoch does not use the acknowledgement feature, and it
is in fact identical to Epoch~1 in Algorithm~{\algGatherOne},
except for a different choice of the parameters. After this epoch, lasting time $O(n)$, all
rumors will be collected in the subtree $\calT'$ consisting of the heavy nodes.

In the second epoch, only the heavy nodes in $\calT'$ will participate in the computation, and
the objective of this epoch is to move all rumors already collected
in $\calT'$ to the root $r$.
The key obstacle to be overcome in this epoch is congestion stemming from the fact that,
although $\calT'$ may be small, its nodes have many rumors to transmit.
This congestion means that simply repeatedly applying $k$-selectors is no longer enough.
For example, if the root has $\ell$ children, each with $n/\ell$ rumors, then repeating an
$\ell$-selector $n/\ell$
times would get all the rumors to the root.  However, we know from the selector size bounds in \cite{Clementi_etal_01} that this would take total time $\Omega(n\ell \log n/\log \ell)$, which is far
too slow. While in this particular scenario $\RoundRobin$ will collect all rumors in the root
in time $O(n)$, this example can be enhanced to show that simple combinations of
$k$-selectors and $\RoundRobin$ do not seem to be sufficient to gather all rumors from $\calT'$
in the root in linear time.

To overcome this obstacle, we introduce two novel tools that will play a critical role
in our algorithm.  The first tool is a so-called \emph{amortizing selector family}.
Since a parent, say with $\ell$ children, receives at most one rumor per round,
it clearly cannot simultaneously be receiving rumors at an average
rate greater than $\frac{1}{\ell}$ from each child individually.
With the amortizing family, we will be able to achieve this bound within a constant fraction
over long time intervals, so long as each child knows (approximately) how many siblings it is competing with.

Of course, such a family will not be useful unless a node can obtain an accurate estimate of its parent's degree,
which will be the focus of our second tool, \emph{$k$-distinguishers}.  Using a
$k$-distinguisher a node can determine whether its number of active siblings $\ell$ is
at least $k$ or at most $2k$.
While this information is not sufficient to determine the exact relation between $\ell$ and $k$,
we show how to combine different $k$-distinguishers to obtain another structure, called a
\emph{cardinality estimator}, that will determine the value of $\ell$ within a factor of $4$.
Using this estimate, and the earlier described amortizing selector, a node
can quickly transmit its rumors to its parent.
This will allow us to gather all rumors from $\calT'$ in the root in time $O(n)$.

This section is divided into three parts.
In Sections~\ref{subsec: amortizing selectors} and~\ref{subsec: k distinguishers}
we give precise definitions and constructions of our combinatorial structures.
Our $O(n)$-time protocol~{\algGatherThree} is described and analyzed
in Section~\ref{subsec: linear time protocol}.


\subsection{Construction of Amortizing Selectors}
\label{subsec: amortizing selectors}

We now define the concept of an \emph{amortizing selector family} $\barS$.
Similarly to a strong selector, this amortizing family will be a collection of subsets of the underlying
label set $[n]$, though now it will be doubly indexed.
Specifically, $\barS = \braced{S_{ij}}$, where
$1 \leq i \leq s$ and each $j \in \braced{1, 2, 4, 8, \dots, k}$,
for some parameters $s$ and $k$.
We say that $\barS$ \emph{succeeds at cumulative rate $q$} if the following statement is true:

For each $j \in \{1, 2, 4, \dots, \frac{k}{2}\}$, each subset $A \subseteq\{1,\dots,n\}$ satisfying $j/2 \leq |A| \leq 2j$, and each element $v \in A$ there are at least $\frac{q}{|A|}s$ distinct $i$ for which
\vskip -0.1in
\begin{equation*}
v \in S_{ij} \textrm{ and } A \cap (S_{i(j/2)} \cup S_{ij} \cup S_{i(2j)}) = \{v\}.
\end{equation*}
In the case $j=1$ the set $S_{i(j/2)}$ is defined to be empty.  Here $s$ can be thought of as the total running time of the selector,  $j$ as a node's estimate of its parent's degree, and $k$ as some bound on the maximum degree handled by the selector.  A node fires at time step $i$ if and only if its index is contained in the set $S_{ij}$.  What the above statement is then saying that for any subset $A$ of siblings, if $|A|$ is at most $k/2$ and each child estimates $|A|$ within a factor of $2$ then each child will transmit at rate at least $\frac{q}{|A|}$.


\begin{theorem} \label{thm:amortizing selector}
There are fixed constants $c,C>0$ such that the following is true: For any $k$ and $n$ and any $s \geq Ck^2 \log n$, there is an amortizing selector with parameters $n,k,s$ succeeding with cumulative rate $c$.
\end{theorem}

Let $k,n,s$ be given such that $s \geq C k^2 \log n$, where $k$ is a constant to be determined later.  We form our selector probabilistically: For each $v,i,$ and $j$, we independently include $v$ in $S_{ij}$ with probability $2^{-j}$.

Observe that by monotonicity it suffices to check the selector property for the case $|A|=2j$: If we replace $A$ with a larger set $A'$ containing $A$ and satisfying $|A'| \leq 4|A|$, then for any $v \in A$ and any $i$ satisfying
$$A' \cap (S_{i(j/2)} \cup S_{ij} \cup S_{i(2j)}) = \{v\},$$
we also have
$$A \cap (S_{i(j/2)} \cup S_{ij} \cup S_{i(2j)}) = \{v\}.$$
So if there are at least $\frac{cs}{|A'|}$ distinct $i$ satisfying the first equality, there are at least $\frac{(c/4)s}{|A|}$ satisfying the second equality.

Now fix $j \in \{1, 2, 4, \dots, k\}$, a set $A\subseteq [n]$ with $|A| = 2j $ and some $v\in A$, and let the random variables $X$ and $Y$ be defined by

\begin{equation*}
	X = |\left\{i: v \in S_{ij} \textrm{ and } A \cap (S_{i(j/2)} \cup S_{ij} \cup S_{i(2j)}) = \{v\} \right\}|
\end{equation*}
The expected value of $X$ is
\begin{equation*}
	\mu_X = s \left(\frac{1}{j}\right) \left(1 - \frac{2}{j}\right)^{2j-1} \left(1 - \frac{1}{j}\right)^{2j-1} \left(1-\frac{1}{2j}\right)^{2j-1}  .
\end{equation*}
Utilizing the bound
$$\left(1 - \frac{2}{j}\right)^{2j-1} \left(1 - \frac{1}{j}\right)^{2j-1} \left(1-\frac{1}{2j}\right)^{2j-1} \geq \frac{1}{e^7},$$
we have
$$\mu_X \geq e^{-7} \frac{s}{j} $$
Now let $c=\frac{1}{4e^7}$. Applying the Chernoff bound, we get
\begin{align*}
		\Prob[ X \le c\frac{s}{|A|} ] \le \Prob[ X \le \frac{1}{2}\mu_X]
						\,&\le \, \e^{-\mu_X/8}	
						\\
						&\le \, \e^{-s/8e^7j}.
\end{align*}

We now use the union bound over all choices of $j$, $v$, and $S$. We have at most $n$ choices of $v$, at most $\log n$ choices for $j$, and at most
$\binom{n}{2j-1}\le n^{2k-1}$ choices of $S$ given $j$ and $v$.   Thus the probability that our family is not an amortizing selector is
at most
\begin{equation*}
	n \log n \cdot n^{2k-1} \cdot \left(e^{-s/8e^7j}\right)
			\le 2 n^{2k} \log n \cdot \e^{-C k \log n / 8e^7}
\end{equation*}
which is smaller than $1$ for sufficiently large $C$.
This implies the existence of the amortized selector family.


\subsection{Construction of $k$-Distinguishers and Cardinality Estimators}
\label{subsec: k distinguishers}

In this section, we define $k$-distinguishers and show how to construct them.
Let $\barS = (S_1,S_2,...,S_m)$, where $S_j \subseteq [n]$ for each $j$.
For $A\subseteq [n]$ and $a\in A$, define
$\Hits{a}{A}{\barS} = \braced{j \suchthat S_j\cap A = \braced{a} }$,
that is
$\Hits{a}{A}{\barS}$ is the collection of indices $j$ for which $S_j$ intersects $A$ exactly on $a$.
Note that, using this terminology, $\barS$ is a strong $k$-selector if and only if
$\Hits{a}{A}{\barS} \neq\emptyset$ for all sets $A\subseteq [n]$ of cardinality at most $k$
and all $a\in A$.

We say that $\barS$ is a \emph{$k$-distinguisher} if there is a \emph{threshold} value $\xi$
(which may depend on $k$) such that, for any $A\subseteq [n]$ and $a\in A$,
the following conditions hold:
\vskip -0.15in
\begin{center}
	if $|A| \le k$ then $|\Hits{a}{A}{\barS}| > \xi$, and if $|A|\ge 2k$ then $|\Hits{a}{A}{\barS}| < \xi$.
\end{center}
\vskip -0.05in
We make no assumptions on what happens for $|A|\in \braced{k+1, k+2,...,2k-1}$.

The idea is this: consider a fixed $a$, and imagine that we have some set $A$ that contains $a$,
but its other elements are not known.
Suppose that we also have an access to a \emph{hit oracle} that for any set
$S$ will tell us whether $S\cap A = \braced{a}$ or not. With this oracle,
we can then use a $k$-distinguisher $\barS$ to
extract some information about the cardinality of $A$ by
calculating the cardinality of $\Hits{a}{A}{\barS}$.
If $|\Hits{a}{A}{\barS}| \le \xi$ then we know that $|A| > k$, and if
$|\Hits{a}{A}{\barS}| \ge \xi$ then we know that $|A| < 2k$.  What we will soon show, again by a probabilistic construction, is that not-too-large $k$-distinguishers exist:


\begin{theorem}\label{thm: distinguisher}
For any $n\ge 2$ and $1\le k \le n/2$
there exists a $k$-distinguisher of size $m =  O(k^2\log n)$.
\end{theorem}

In our framework, the acknowledgement of a message received from a parent corresponds exactly to such a hit oracle.  So if all nodes fire according to such a $k$-distinguisher, each node can determine in time $O(k^2 \log n)$ either that its parent has at least $k$ children or that it has at most $2k$ children.

Now let $\lambda$ be a fixed parameter between $0$ and $1$.  For each $i = 0,1,...,\ceiling{\lambda\log n}$, let $\barS^i$ be a $2^i$-distinguisher of
size $O(2^{2i}\log n)$ and with threshold value $\xi_i$.
We can then concatenate these $k$-distinguishers to obtain a sequence
$\tildeS$ of size
$\sum_{i=0}^{\ceiling{\lambda\log n}}O(2^{2i}\log n)  = O(n^{2\lambda}\log n)$.

We will refer to $\tildeS$ as a \emph{cardinality estimator}, because applying our hit oracle to $\barS$
we can estimate a cardinality of an unknown set within a factor of $4$,
making $O(n^{2\lambda}\log n)$ hit queries.   More specifically, consider again a scenario where we have a fixed $a$ and some unknown set $A$ containing $a$, where $|A|\le n^{\lambda}$.  Using the hit oracle, compute the values $h_i = |\Hits{a}{A}{\barS^i}|$, for all $i$.  If $i_0$ is the smallest $i$ for which $h_i > \xi_i$, then by definition of our distinguisher we must have
$2^{i_0-1} < |A| < 2( 2^{i_0})$.
In our gathering framework, this corresponds to each node in the tree being able to determine in time $O(n^{2 \lambda} \log n)$ a value of $j$ (specifically, $i_0-1$) such that the number of children of its parent is between $2^j$ and $2^{j+2}$, which is exactly what we need to be able to run the amortizing selector.

It remains to show the existence of $k$-distinguishers, i.e. to prove Theorem \ref{thm: distinguisher}.  Let $m= C k^2 \log n$, where $C$ is some sufficiently large constant whose value we will
determine later.  We choose the
collection of random sets $\barS = (S_1,S_2,...,S_m)$, by letting each $S_j$ be formed by independently including
each $x\in [n]$ in $S_j$ with probability $1/2k$.
Thus, for any set $A$ and $a\in A$,
the probability that $S_j\cap A = \braced{a}$ is $(1/2k)(1-1/2k)^{|A|-1}$, and the
expected value of $|\Hits{a}{A}{\barS}|$ is
\begin{equation}
\textstyle
{\hbox{\bf E}}[|\Hits{a}{A}{\barS}|] =	m \cdot \frac{1}{2k}\left(1-\frac{1}{2k}\right)^{|A|-1}.
	\label{eq: exp number hits}
\end{equation}

Recall that to be a $k$-distinguisher our set needs to satisfy (for suitable $\xi$) the following two properties:
$$\begin{tabular}{ll}
	(d1) & if $|A| \le k$ then $|\Hits{a}{A}{\barS}| > \xi$ \\
(d2) & if $|A|\ge 2k$ then $|\Hits{a}{A}{\barS}| < \xi$. \end{tabular}$$

We claim that, for a suitable value of $\xi$,
the probability that there exists a set $A\subseteq [n]$ and
some $a\in A$ for which $\barS$ does not satisfy
both conditions is smaller than $1$ (and in fact tends to $0$) This will be sufficient to show
an existence of a $k$-distinguisher with threshold value $\xi$.

\smallskip

Observe that in order to be a $k$-distinguisher it is sufficient that
$\barS$ satisfies (d1) for sets $A$ with $|A| = k$ and satisfies (d2)
for sets $A$ with $|A| = 2k$. This is true because the value of
$|\Hits{a}{A}{\barS}|$ is monotone with respect to the inclusion:
if $a \in A\subseteq B$ then $\Hits{a}{A}{\barS}\supseteq \Hits{a}{B}{\barS}$.

\smallskip

Now consider some fixed $a\in[n]$ and
two sets $A_1, A_2 \subseteq [n]$ such that $|A_1| = k$, $|A_2| = 2k$ and $a\in A_1\cap A_2$.
For $i =1,2$, we consider two corresponding random variables $X_i = |\Hits{a}{A_i}{\barS}|$ and
their expected values $\mu_i = \Exp[X_i]$. For any integer $k\ge 1$ we have
\begin{align*}
\textstyle
\frac{1}{\e^{1/2}}  &\le \textstyle (1-\frac{1}{2k})^{k-1}
\\
\textstyle
\frac{1}{\e} &\le \textstyle	(1-\frac{1}{2k})^{2k-1} \le \half.
\end{align*}
From~(\ref{eq: exp number hits}), substituting $m = Ck^2\log n$,
this gives us the corresponding estimates for $\mu_1$ and $\mu_2$:
\begin{align*}
\textstyle
\frac{1}{2e^{1/2}} C k \log n &\le  \mu_1
	\\
\textstyle
\frac{1}{2\e} C k \log n	&\le	\mu_{2} \le \onefourth C k \log n
\end{align*}
Since $\e^{-1/2} > \half$, we can choose an $\epsilon \in (0,1)$ and $\xi$ for which
\begin{equation*}
	(1+\epsilon)\mu_2  < \xi < (1-\epsilon)\mu_1.
\end{equation*}
Thus the probability that $\barS$ violates (d1) for $A=A_1$ is
\begin{align*}
	\Prob[ X_1 \le \xi ] \,\le\, \Prob[ X_1 \le (1-\epsilon)\mu_1]
					\,&\le \, \e^{-\epsilon^2\mu_1/2}	
					\\
					&\le \, \e^{-\epsilon^2 \e^{-1/2} C k \log n/ 4},
\end{align*}
where in the second inequality we use the Chernoff bound for deviations below the mean.
Similarly, using the Chernoff bound for deviations above the mean, we can bound
the probability of $\barS$ violating (d2) for $A=A_2$ as follows:
\begin{align*}
	\Prob[ X_2 \ge \xi ] \,\le\, \Prob[ X_2 \ge (1+\epsilon)\mu_2]
 				\,&\le\, e^{-\epsilon^2\mu_2/3}
				\\
				\le\, e^{- \epsilon^2\e^{-1} C k\log n/6}.
\end{align*}
To finish off the proof, we apply the union bound.
We have at most $n$ choices for $a$,
at most $\binom{n}{k-1}\le n^{k-1} \le n^{2k-1}$ choices of $A_1$, and at most
$\binom{n}{2k-1}\le n^{2k-1}$ choices of $A_2$. Note also that
$\e^{-1/2}/4 > \e^{-1}/6$. Putting it all together, the probability that
$\barS$ is not a $k$-distinguisher is at most
\begin{align*}
\textstyle
n \cdot {\binom{n}{k-1}} \cdot \Prob[ X_1 \le \xi ]  + n \cdot {\binom{n}{2k-1}} \cdot \Prob[ X_2 \ge \xi ]
		&\le  n^{2k} \cdot  (  \Prob[ X_1 \le \xi ]+\Prob[ X_2 \ge \xi ])
		\\
		&\le 2 n^{2k} \cdot e^{- \epsilon^2\e^{-1} C k\log n/6}
		\\
		&= 2n^{k(2 - \epsilon^2\e^{-1} C/6)}
		< 1,
\end{align*}
for $C$ large enough.


\subsection{Linear-time Protocol}
\label{subsec: linear time protocol}

As before, $\calT$ is the input tree with $n$ nodes. We will recycle the notions of light and
heavy nodes from Section~\ref{sec: simple algorithm}, although now we will use
slightly different parameters. Let $\delta>0$ be a small constant, and let
$K = \ceiling{n^\delta}$.
We say that $v\in\calT$ is \emph{light} if $|T_v|\le n/K^3$ and we call $v$
\emph{heavy} otherwise. By $\calT'$ we denote the subtree of $\calT$ induced by the
heavy nodes.

\myparagraph{Algorithm~{\algGatherThree}.}
Our algorithm will consist of two epochs.
The first epoch is essentially identical to Epoch~1 in Algorithm~{\algGatherOne},
except for a different choice of the parameters.
The objective of this epoch is to collect all rumors in $\calT'$ in time $O(n)$.
In the second epoch, only the heavy nodes in $\calT'$ will participate in the computation, and
the objective of this epoch is to gather all rumors from $\calT'$ in the root $r$.
This epoch is quite different from our earlier algorithms and it will use the
combinatorial structures obtained in the previous sections to
move all rumors from $\calT'$ to $r$ in time $O(n)$.


\smallskip

\emparagraph{Epoch~1:} In this epoch only light nodes will participate, and the
objective of Epoch~1 is to move all rumors into $\calT'$.
In this epoch we will not be taking advantage of the acknowledgement mechanism.
As mentioned earlier, except for different choices of parameters, this epoch is
essentially identical to Epoch~1 of Algorithm~{\algGatherOne}, so we only give
a very brief overview here. We use a strong $K$-selector $\barS$ of
size $m\le CK^2\log n$.

Let $D= \lceil \log_K n \rceil \leq 1/\delta=O(1)$.  By Lemma~\ref{lem: gamma-heights}, the $K$-depth of $\calT$ is at most $D$.  Epoch $1$ consists of $D+1$ stages, where in each stage $h = 0,1,...,D$,
nodes of $K$-depth $h$ participate. Stage $h$ consists of $n/K^3$ executions of $\barS$,
followed by an execution of {\RoundRobin}, taking total time $n/K^3 \cdot m +n=O(n)$.  So the entire epoch takes time $(D+1)\cdot O(n) = O(n)$ as well.
The proof of correctness (namely that after this epoch all rumors are in $\calT'$)
is identical as for Algorithm~{\algGatherOne}.


\smallskip

\emparagraph{Epoch~2:} When this epoch starts, all rumors are already
gathered in $\calT'$, and the objective is to push them further to the root.

For the second epoch, we restrict our attention to the tree $\mathcal T'$ of heavy nodes.  As before, no parent in this tree can have more than $K^3=n^{3\delta}$ children, since each child is itself the ancestor of a subtree of size $n/K^3$.  We will further assume the existence of a fixed amortizing selector family with parameters $k=2 K^3$ and $s=K^8$, as well as a fixed cardinality estimator with parameter $\lambda=3 \delta$ running in time $D_1=O\left(n^{6 \delta} \log n\right) = O\left(K^6 \log n\right)$.

Our protocol will be divided into stages, each consisting of $2(D_1+K^8)$ steps.
   A node will be \emph{active} in a given stage if at the beginning of the stage it has already received all of its rumors, but still has at least one rumor left to transmit (it is possible for a node to never become active, if it receives its last rumor and then finishes transmitting before the beginning of the next stage).

During each odd-numbered time step of a stage, all nodes (active or not) holding at least one rumor they have not yet successfully passed on transmit such a rumor.
The even-numbered time steps are themselves divided into two parts.  In the first $D_1$ even steps, all active nodes participate in the aforementioned cardinality estimator.  At the conclusion of the estimator, each node knows a $j$ such that their parent has between $2^j$ and $2^{j+2}$ active children.  Note that active siblings do not necessarily have the same estimate for their family size.  For the remainder of the even steps, each active node fires using the corresponding $2^{j+1}$-selector from the amortizing family.

The stages repeat until all rumors have reached the root.  Our key claim, stated below, is that the rumors aggregate at least at a steady rate over time -- each node with subtree size $m$ in the original tree $\calT$ will have received all $m$ rumors within $O(m)$ steps of the start of the epoch.


\begin{description}
\item{\inv{I}{}{\ell}} For any heavy node $v$ such that $v$ has subtree size $m$ in $\calT$, and any $0 \leq j \leq m$, that node has received at least $j$ rumors within time $C(2m+j)$ of the beginning of Epoch $2$,
where $C$ is some sufficiently large absolute constant.
\end{description}

In particular, the root has received all of the rumors by time $3Cn$.

We will show this invariant holds by induction on the node's height within $\mathcal T'$.  If the node is a leaf the statement follows from our analysis of Epoch $1$ (the node has received \textit{all} rumors from its subtree by the beginning of epoch $2$).

Now assume that a node $u$ with subtree size $m+1$ has $k$ children within $\mathcal T'$, and that those children have subtree sizes $a_1 \geq a_2 \geq \dots \geq a_k \geq K^3$.  Node $u$ may also received some number $a_0$ of messages from non-heavy children (these children, if any, will have already completed their transmissions during the previous epoch).

Let $v$ be a child having subtree size $a_1$ (chosen arbitrarily, if there are two children with maximal subtree size).  Let $t_2$ be defined by
$$t_2 = \left\{ \begin{array}{cc} 3C a_2 + \frac{3}{c} \left(a_2+\dots+a_k\right)+K^{12} & \textrm{ if } k>1 \\ 0 & \textrm{ if } k=1 \end{array}\right.$$
We make the following additional claims.

\smallskip

\noindent \textbf{Claim 2:} By time $t_2$, all children except possibly $v$ will have completed all of their transmissions.

\begin{proof} By inductive hypothesis, all children except $v$ will have received all of their rumors by time $3Ca_2$.  During each stage from that point until all non-$v$ nodes complete, one of the following things must be true.
\begin{itemize}
\item{Some active node transmits the final rumor it is holding, and stops transmitting.  This can happen during at most $K^3$ stages, since there are at most $K^3$ children and each child only becomes active once it already has received all rumors from its subtree.}
\item{All active nodes have rumors to transmit throughout the entire stage.  If there were $j$ active nodes total during the stage, then by the definition of our amortizing selector family, the parent received at least $c \frac{2K^8}{j}$ rumors from each child during the stage.  In particular, it must have received at least
    $$c(j-1) \frac{2K^8}{j} \geq c K^8$$
    new rumors from children other than $v$.  }
\end{itemize}
Combining the two types of stages, the non-$v$ children will have all finished in at most
$$K^3+\frac{1}{c K^8} \left(a_2+\dots+a_k\right)$$
complete stages after time $3Ca_2$.  Since each stage takes time $2(D_1+K^8) = (2+o(1))K^8$, the bound follows.
\end{proof}

\noindent \textbf{Claim 3:} Let $k, m,$ and $v$ be as above.  By time $2Cm$, all children except possibly $v$ have completed their transmissions.

\begin{proof}
This is trivial for $k=1$.  For larger $k$ follows from the previous claim, together with the estimate that (for sufficiently large $C$)
$$2Cm \geq 2C(a_1+\dots+a_k) \geq 4C a_2 + 2C (a_3+\dots+a_k) \geq \frac{4}{3} (t_2-K^{12}) \geq \frac{5}{4} t_2$$
Here the middle inequality holds for any $C>2/c$, while the latter inequality holds since $t_2 \geq a_2 \geq n/K^3 \gg K^{12}$.  \end{proof}

By the above claim, node $v$ is the only node that could still be transmitting at time $2Cm$.  In particular, if it has a rumor during an odd numbered time step after this point, it successfully transmits.  By assumption, $v$ will have received at least $j$ rumors by time $C(2m+j)$ for each $j$.  This implies it will successfully broadcast $j$ rumors by time $C(2m+j)+2$ for each $0 \leq j \leq a_1$.

By time $C[2(m+1)+j]$, the parent has either received all rumors from all of its children (if $j>a_1$), or at least $j$ rumors from $a_1$ alone (if $j \leq a_1$).  Either way, it has received at least $j$ rumors total, and the induction is complete.


\bibliographystyle{plain}
\bibliography{gather_trees}

\end{document}